\documentclass[12pt]{article}
\title{\bf Three-Dimensional Ising Model and Transfer Matrices
\rm}

\author{S.L.Lou\thanks{~E-mail: {\tt sllou@mail.thu.edu.tw}}\,\,and\,\,S.H.Wu \\ Department of physics, Tunghai
University, \\ Taichung, Taiwan407, R.O.C.}

\begin{document}

\begin{titlepage}
\maketitle \vspace{3cm} \bf \abstract{\rm The use of a transfer
matrix method to solve the 3D Ising model is   straightforwardly
generalized from the 2D case. We follow  B.Kaufman's approach. No
approximation is made, however  the largest eigenvalue cannot be
identified. This problem comes from the fact that we follow the
choice of directions of 2-dimensional rotations in the direct
product space of the 2D Ising model such that all eigenvalue
equations reduce miraculously to only one equation.
 Other choices of directions of 2-dimensional rotations for finding
the largest eigenvalue may lose this fascinating feature.
Comparing the series expansion of internal energy per site at the
high temperature limit with the series obtained from the computer
graphic method, we find these two series have very similar
structures. A possible correct via a factor $\Phi(x)$ is suggested
to fit the result of the graphic method.}

\vspace{1cm} \noindent PACS:05.50.+q
\end{titlepage}
\addtocounter{page}{1}
\section {Introduction} Although over a half-century has passed,
solving the 3D Ising model exactly is still an open problem.
Anyone who claims to solve this model exactly should, at least,
evaluate its partition function, internal energy per site,
critical temperature and the critical exponents $ \alpha $ and $
\beta$ calculated from the relevant heat capacity and
magnetization per site individually. In addition, a crucial test,
similar to one L.Onsager \cite{On} did in 1944, to check whether
the results are right or wrong, is that one should compare the
series expansion coefficients of the internal energy per site at
the high temperature limit with the series obtained from some
other methods \cite{IT} e.g., the computer graphic method, at
least up to the first three or four nonvanishing terms \cite{DG}
  \cite{ZJ} \cite{LF} \cite{GE} \cite{RE} \cite{NM}.

Among the many various methods for deriving the partition function
of the 2D Ising model, transfer matrix method is the oldest and
original method. However, the generalization of this method to the
3D case has had relatively little discussion. In this paper, we
have no ambition to solve this 3D Ising model  satisfying all of
the requirements mentioned above. Instead, B.Kaufman's approach
\cite{Ka} in the 3D Ising model is carried out step by step. Any
approximation is avoided if we possibly  can. In the following, it
is shown that, when a transfer matrix formalism is set up, a
spinor representation can work. 2-dimensional rotations in the
direct product space and the feature that all of the eigenvalue
equations reduce miraculously to only one equation also appear in
the 3D Ising model. Even though the final high-temperature
expansion series of internal energy per site is not exactly the
same as the computer graphic method's, these two series do have
the same structures. This discrepancy may be related to a dilemma
between the choice of the directions of the 2-dimensional
rotations in order to find the largest eigenvalue and losing the
fascinating feature that all of the eigenvalue equations reduce to
only one equation. Be it ever not so perfect, we hope this
generalization may lay the foundations for further study.
\section{Transfer Matrices}

Let us consider a simple cubic lattice with $ l$ layers, each has
$ m $ rows and $n$ sites per row. So there are N points on the
lattice, N=$mnl$. Periodic boundary conditions are used. To each
lattice point, with integral coordinates $\tau$, $\rho$, $\zeta$,
we assign a spin variable s($\tau$,$\rho$,$\zeta$) which takes two
values $\pm$ 1. The energy of the configuration is given by
\begin{equation}
E (s)=-J\sum_{\tau =1}^n \sum_{\rho =1}^m \sum_{\zeta =1}^l \{
s(\tau , \rho , \zeta ) s(\tau +1, \rho , \zeta )+ s(\tau , \rho,
\zeta ) s(\tau , \rho +1, \zeta )+s(\tau , \rho , \zeta ) s(\tau ,
 \rho , \zeta +1)\} .
\end{equation}
J($>$0) is the coupling of a pair of neighboring spins. The
partition function
\begin{eqnarray}
\mathcal{Z} &=& \sum_{(s)} e^{-E(s) / T} \\ &=& \sum_{( s )}
\prod^n_{\tau=1} \prod^m_{\rho=1} \prod^l_{\zeta=1} e^{Ks(\tau ,
\rho , \zeta ,) s(\tau +1, \rho , \zeta )} e^{Ks(\tau , \rho ,
\zeta ) s(\tau , \rho +1 , \zeta )} e^{Ks(\tau , \rho , \zeta )
s(\tau ,\rho ,\zeta +1) },
\end{eqnarray}
is taken over all the $2^N$ possible configurations. Here $
K\equiv J/T$. Now we factor the partition function into terms each
involving only two neighboring spins, giving

\begin{eqnarray}
\mathcal{Z} &=& \sum_{s(1, \cdot ,\cdot)} \cdots \sum_{s(n, \cdot
,\cdot )}  \big< s (1, \cdot ,\cdot) \mid \mathcal{V} \mid s(2,
\cdot ,\cdot )\big> \big< s(2, \cdot ,\cdot ) \mid \mathcal{V}
\mid s(3,\cdot ,\cdot )\big> \cdots \\ & & \; \; \; \; \; \; \; \;
\; \; \; \; \; \; \; \; \; \; \; \; \; \big< s(n, \cdot ,\cdot
)\mid \mathcal{V} \mid s(1, \cdot ,\cdot )\big> \nonumber \\
\nonumber
\\ &=& Tr \mathcal{V}^n ,
\end{eqnarray}
where the matrix elements of the transfer matrix $\mathcal{V}$ are
\begin{equation}
\big< s(\tau ,\cdot ,\cdot ) \mid \mathcal{V} \mid s(\tau +1,
\cdot ,\cdot ) \big> = \prod_{\rho =1}^m \prod_{\zeta =1}^l
e^{Ks(\tau ,\rho ,\zeta )s(\tau +1, \rho ,\zeta )} e^{Ks(\tau
,\rho ,\zeta )s(\tau ,\rho +1, \zeta )} e^{Ks(\tau ,\rho ,\zeta
)s(\tau ,\rho ,\zeta +1) }.
\end{equation}
$\mathcal{V}$ can be put into a more convenient form by factoring
it into the product of simpler matrices,
\begin{eqnarray}
\mathcal{V} &=&  \mathcal{V}_3 \mathcal{V}_2 \mathcal{V}_1 ,
\\ \nonumber
\\
 \big< s( \tau ,\cdot ,\cdot ) \mid \mathcal{V}_1
 \mid s(\tau +1 \cdot ,\cdot )\big>
 &=& \prod_{\rho = 1}^m \prod_{\zeta = 1}^l
 e^{Ks(\tau ,\rho ,\zeta )s(\tau +1,\rho , \zeta )} ,\\ \nonumber \\
 \big< s(\tau ,\cdot ,\cdot ) \mid \mathcal{V}_2
 \mid s(\tau +1 \cdot ,\cdot )\big>
 &=& \prod_{\rho =1}^m \prod_{\zeta =1}^l
 e^{Ks(\tau , \rho ,\zeta )s(\tau ,\rho +1, \zeta )  }
\; \delta_{s(\tau ,\rho ,\zeta )s(\tau +1 ,\rho ,\zeta )}, \\
\nonumber \\
 \big< s(\tau ,\cdot ,\cdot ) \mid \mathcal{V}_3 \mid
 s(\tau +1, \cdot ,\cdot ) \big>
 &=& \prod_{\rho =1}^m \prod_{\zeta =1}^l
 e^{Ks(\tau ,\rho ,\zeta )s(\tau ,\rho ,\zeta +1)}
\; \delta_{s(\tau ,\rho ,\zeta )s(\tau +1, \rho ,\zeta )}.
\end{eqnarray}
 The above decomposition may be checked as follows:
\begin{eqnarray}
&& \big< s(\tau ,\cdot ,\cdot) \mid \mathcal{V}_3 \mathcal{V}_2
\mathcal{V}_1 \mid s(\tau +1 ,\cdot ,\cdot ) \big> \\ \nonumber \\
&=& \sum_{s(\tau +1 ,\cdot ,\cdot )} \sum_{s'(\tau , \cdot ,\cdot
)} \sum_{s'(\tau +1, \cdot ,\cdot )} \sum_{s^{''}(\tau ,\cdot
,\cdot )} \sum_{s^{''}(\tau +1, \cdot ,\cdot )} \big< s(\tau
,\cdot ,\cdot ) \mid \mathcal{V}_3 \mid s(\tau +1,\cdot ,\cdot )
\big> \nonumber
\\ && \; \; \; \; \; \; \; \; \; \; \; \; \; \; \;
\big< s(\tau +1,\cdot ,\cdot) \mid s'(\tau,\cdot,\cdot) \big>
\big< s'(\tau , \cdot,\cdot) \mid \mathcal{V}_2 \mid s'(\tau
+1,\cdot,\cdot) \big> \nonumber
\\ && \; \; \; \; \; \; \; \; \; \; \; \; \; \; \;
\big< s'(\tau +1,\cdot,\cdot )\mid s^{''}(\tau,\cdot,\cdot)\big>
\big< s^{''}(\tau ,\cdot,\cdot ) \mid \mathcal{V}_1 \mid s^{''}
(\tau +1, \cdot ,\cdot ) \big> \nonumber \\  && \; \; \; \; \; \;
\; \; \; \; \; \; \; \; \;
 \big< s^{''}(\tau +1
,\cdot ,\cdot ) \mid s(\tau +1,\cdot ,\cdot ) \big> \\ \nonumber
\\ &=& \prod_{\rho =1}^m \prod_{\zeta=1}^l \{  \sum_{s(\tau
+1,\rho ,\zeta )} \sum_{s'(\tau ,\rho ,\zeta )} \sum_{s'(\tau +1,
\rho ,\zeta )} \sum_{s^{''}(\tau ,\rho ,\zeta )} \sum_{s^{''}
(\tau +1,\rho ,\zeta )} \nonumber \\  & & \; \; \; \; \; \; \; \;
\; \; \; \; \; \; \;
 \big( e^{Ks(\tau ,\rho ,\zeta)s(\tau ,\rho
,\zeta +1)} \delta_{s(\tau ,\rho ,\zeta )s(\tau +1,\rho ,\zeta )}
\big) \big( \delta_{s(\tau +1,\rho ,\zeta )s'(\tau ,\rho ,\zeta )}
\big) \nonumber \\ && \; \; \; \; \; \; \; \; \; \; \; \; \; \; \;
 \big( e^{Ks'(\tau ,\rho ,\zeta)s'(\tau ,\rho +1 ,\zeta )}
\delta_{s'(\tau ,\rho ,\zeta )s'(\tau +1,\rho ,\zeta )} \big)
 \big( \delta_{s'(\tau +1,\rho ,\zeta )s^{''}(\tau ,\rho ,\zeta
)} \big) \nonumber \\ && \; \; \; \; \; \; \; \; \; \; \; \; \; \;
\;
 \big( e^{Ks^{''}(\tau ,\rho ,\zeta )s^{''}(\tau +1, \rho
,\zeta )} \big)  \big( \delta_{s^{''}(\tau +1, \rho ,\zeta )s(\tau
+1,\rho ,\zeta )} \big) \}  \\ \nonumber \\ &=& \prod_{\rho =1}^m
\prod_{\zeta =1}^l e^{Ks(\tau ,\rho ,\zeta )s(\tau +1,\rho
,\zeta)} e^{Ks(\tau ,\rho ,\zeta )s(\tau ,\rho +1 ,\zeta)}
e^{Ks(\tau ,\rho ,\zeta )s(\tau ,\rho ,\zeta +1)}.
\end{eqnarray}
In the above equation, due to periodic boundary conditions, the
identity
\begin{equation}
\prod_{\rho = 1}^m \prod_{\zeta =1}^l \delta_{s(\tau ,\rho , \zeta
)s'(\tau ,\rho ,\zeta ) }
 = \prod_{\rho =1 }^m \prod_{\zeta =1 }^l
 \delta_{s(\tau ,\rho +1, \zeta )s'(\tau ,\rho +1 ,\zeta )}
\end{equation}
is used. Furthermore, $\mathcal{V}_1 , \mathcal{V}_2 ,
\mathcal{V}_3 $ can be rewritten as a matrix in the direct product
space. Observing from (8), let us define a matrix $a$ with matrix
elements
\begin{equation}
\big< s(\tau ,\rho ,\zeta ) \mid a \mid s(\tau +1 ,\rho , \zeta )
\big> = e^{Ks(\tau ,\rho ,\zeta )s(\tau +1, \rho ,\zeta )}.
\end{equation}
\begin{eqnarray}
a &=& \left( \begin{array}{cc} e^K & e^{-K}
\\ e^{-K} & e^K \end{array} \right) \\ \nonumber \\
&=& e^K \big( I + e^{-2K} \sigma_x \big) \\ \nonumber \\ &=& \big(
2\; sinh2K \big)^{1/2} e^{K^* \sigma_x }.
\end{eqnarray}
$I$ is a $2\times 2$ unit matrix. $tanh K^* = e^{-2K} ,$ $tanh K =
e^{-2K^*} ,$ $sinh 2K \; sinh 2K^* =1 .$
 To simplify the matrics $\mathcal{V}_1 ,\mathcal{V}_2 ,
 \mathcal{V}_3 $, we define
 \begin{eqnarray}
 X_{i,j} &=& \underbrace{\overbrace{\big( I\otimes I\otimes \cdots
 \otimes I \big)}^{m\; times \; of \; I}}_{\zeta =1} \otimes \cdots \otimes
 \underbrace{\big( I\otimes \cdots \otimes \overbrace{\sigma_x}^
 {\rho =i}
 \otimes \cdots \otimes I \big) }_{\zeta =j } \cdots \nonumber \\
  \nonumber \\
  & & \otimes \underbrace{ \big( I\otimes I \otimes \cdots
  \otimes I \big) }_{\zeta =l }.
 \end{eqnarray}
 $Y_{i,j} $ and $Z_{i,j}$ are also defined similarly by replacing
 the Pauli matrix $\sigma_x $ with $\sigma_y$ and $\sigma_z$
 respectively.
 \begin{eqnarray}
 \mathcal{V}_1 &=& \overbrace{a\otimes a\otimes \cdots \otimes a}^{
 ml \; times} \\ \nonumber \\
 &=& \big( 2 sinh 2K \big)^{ml/2} \prod_{\rho =1}^m \prod_{\zeta
 =1}^l e^{K^* X_{\rho ,\zeta }}.
 \end{eqnarray}
 As for $\mathcal{V}_2$, we introduce  another matrix $b$ with matrix
 elements
 \begin{equation}
 \big< s(\tau ,\rho ,\zeta ) \mid  b \mid
 s(\tau +1 ,\rho ,\zeta ) \big>
 = \delta_{s(\tau ,\rho ,\zeta ) s(\tau +1 ,\rho ,\zeta ) }
 e^{Ks(\tau ,\rho ,\zeta ) s(\tau ,\rho +1 ,\zeta )},
 \end{equation}
 \begin{eqnarray}
 b &=& \left(  \begin{array}{llll} e^K & 0 & 0 & 0 \\ 0 & e^{-K} & 0 &
 0 \\0 & 0 & e^K & 0 \\ 0 & 0 & 0 & e^{-K} \end{array}
 \right)=e^{K\sigma _z \otimes \sigma_z} =e^{K(\sigma_z \otimes
 I)(I\otimes \sigma_z)} ,\\ \nonumber \\
 \mathcal{V}_2 &=& \overbrace{b \otimes b \otimes \cdots
 \otimes b }^{ml \; times} \\ \nonumber \\ &=&
  \prod_{\rho=1} ^m \prod_{\zeta=1}^l e^{KZ_{\rho+1 ,\zeta} Z_{\rho
  ,\zeta }}.
 \end{eqnarray}
 Similarly, $\mathcal{V}_3$ is obtained from a
 matrix c,
 \begin{eqnarray}
 c &=& e^{K(\sigma _z \overbrace{\otimes I \otimes \cdots \otimes
 I}^{m \; times })\;(\overbrace{ I\otimes I\otimes \cdots I \otimes }^{m\;
 times }  \sigma _z )}.
 \\ \nonumber \\ \mathcal{V}_3 &=& \overbrace{c \otimes c\otimes \cdots \otimes
 c}^{ml \; times} \\ \nonumber \\ &=& \prod_{\rho =1}^m \prod_{\zeta=1}^l
 e^{KZ_{\rho ,\zeta}Z_{\rho ,\zeta +1}}\; .
 \end{eqnarray}
 \section{Spinor Representation}
 $\mathcal{V}_1$, $\mathcal{V}_2$,$\mathcal{V}_3$
  in $2^{ml}$-space can be related to matrices in
 2$ml$ -spaces via Dirac $\Gamma$ matrices. The process of
 reducing
 the dimensions of $\mathcal{V} $  had been used in the 2D Ising model.
  Define a set of matrix $\Gamma_{\mu ,\zeta}$
 satisfying anticommutation relations
 \begin{equation}
 \Gamma_{\mu ,\zeta} \Gamma _{\mu^{'}, \zeta^{'}}+\Gamma_{\mu^{'},
 \zeta^{'}} \Gamma_{\mu, \zeta} =2\delta_{\mu \mu^{'}}
 \delta_{\zeta \zeta^{'}},
 \end{equation}

 \begin{equation}
 (\mu, \mu^{'}=1,2 \cdots 2m; \; \;  \zeta, \zeta^{'}=1,2 \cdots l
 ). \nonumber
 \end{equation}
 Every $\Gamma_{\mu, \zeta}$ is a $2^{ml}$ $\times$ $2^{ml}$
 matrix. A possible representation of $\Gamma_{\mu ,\zeta}$ is
 \begin{eqnarray}
 \Gamma_{1,1} &=& Z_{1,1} ,
 \\ \Gamma_{2,1} &=& Y_{1,1} ,
 \\ \Gamma_{3,1} &=& X_{1,1}Z_{2,1} ,
 \\ \Gamma_{4,1} &=& X_{1,1}Y_{2,1} ,
 \\ \vdots \nonumber
 \\ \Gamma_{2m-1,1} &=& X_{1,1} X_{2,1}\cdots X_{m-1,1}Z_{m,1} ,
 \\ \Gamma_{2m,1} &=& X_{1,1}X_{2,1}\cdots X_{m-1,1} Y_{m,1} ,
 \\ \Gamma_{1,2} &=& \big( \; X_{1,1}X_{2,1}\cdots X_{m,1} ,
 \;\big)\;Z_{1,2}=U_1 Z_{1,2} ,
 \\ \Gamma_{2,2} &=& \big(\; X_{1,1}X_{2,1}\cdots X_{m,1}
 \;\big)\; Y_{1,2}=U_1 Y_{1,2} ,
 \\ \vdots \nonumber
 \\ \Gamma_{1,\zeta} &=& U_1U_2\cdots U_{\zeta -1}Z_{1,\zeta} ,
 \\ \Gamma_{2,\zeta} &=& U_1U_2\cdots U_{\zeta -1}Y_{1,\zeta} ,
 \\ \vdots \nonumber
 \\ \Gamma_{2m-1, l} &=& U_1U_2\cdots U_{l-1}Z_{m, l} ,
 \\ \Gamma_{2m,l} &=& U_1U_2\cdots U_{l-1}Y_{m, l} ,
 \end{eqnarray}
 where
 \begin{equation}
 U_{\zeta}\equiv \big(I\otimes \cdots \otimes I \big) \otimes \cdots
 \otimes \big( \underbrace{ \sigma_x \otimes \sigma_x \otimes \cdots \otimes
 \sigma_x}_{\zeta^{th}\; block} \big) \otimes \big( I \otimes \cdots \otimes I \big)
 \cdots \otimes \big(I \otimes \cdots \otimes I \big) .
 \end{equation}
 The total number of the $\Gamma$ matrix is 2$ml$. A special $2^{ml}
 \times 2^{ml}$ matrix U is defined as
 \begin{eqnarray}
 U &=& \underbrace{\sigma_x \otimes \sigma_x \otimes \cdots \otimes \sigma
 _x}_{ml\; times} \\ \nonumber \\ &=& i^{ml} \prod_{\zeta=1}^{l}\; \Big(\;
 \prod_{\mu=1}^{2m}\;
 \Gamma_{\mu, \zeta}\;\Big).
 \end{eqnarray}
 U and $U_{\zeta}$ have the following relations:
 \begin{equation}
 U^2=I  ,\;\;\;\;\;\;\;\;\;\; U_{\zeta}^2=I,
 \end{equation}
 \begin{equation}
 \; \;\;  U\big(I+U \big)=I+U ,\;\;\;\;\;\;\;\;\;\;\; U_{\zeta}
 \big(I+U_{\zeta} \big)=I+U_{\zeta},
 \end{equation}
 \begin{equation}
 \; \; \;  U \big( I-U \big)=U-I ,\;\;\;\;\;\;\;\;\;\; U_{\zeta}
 \big( I-U_{\zeta} \big)=U_{\zeta} -I,
 \end{equation}
 \begin{equation}
  \{ U, \Gamma_{\mu, \zeta} \}=0 ,\;\;\;\;\;\;\;\;\;\;
 \{U_{\zeta}, \Gamma_{\mu, \zeta} \} =0,
 \end{equation}
 \begin{equation}
 \big[ U, \Gamma_{\mu, \zeta} \Gamma_{\mu^{'}, \zeta^{'}} \big] =0
 ,\;\;\;\;\;\;\;\;\;\; \big[ U_{\zeta}, \Gamma_{\mu, \zeta}
 \Gamma_{\mu^{'}, \zeta } \big]=0.
 \end{equation}
 By definition of $\Gamma$, we notice that
 \begin{equation}
 \Gamma_{2\rho, \zeta} \Gamma_{2\rho -1, \zeta}=Y_{\rho,\zeta}
 Z_{\rho, \zeta}=iX_{\rho, \zeta}\;,
 \end{equation}
 then
 \begin{equation}
 \mathcal{V}_1 = \big( 2 sinh2K \big)^{ml/2}
\prod_{\rho=1}^m\prod_{\zeta=1}^l e^{i K^* \Gamma_{2\rho-1,
 \zeta}\Gamma_{2\rho, \zeta}}\; .
 \end{equation}
 Simlarly,
 \begin{equation}
 \Gamma_{2\rho+1, \zeta} \Gamma_{2\rho, \zeta}=iZ_{\rho,
 \zeta}Z_{\rho+1, \zeta}\; ,
 \end{equation}
 then we have
 \begin{eqnarray}
 \mathcal{V}_2 &=& \prod_{\zeta=1}^l \{ e^{KZ_{m,
 \zeta}Z_{1,\zeta}} \prod_{\rho=1}^{m-1}
 e^{KZ_{\rho,\zeta}Z_{\rho+1, \zeta}} \}
 \\ \nonumber \\ &=& \prod_{\zeta=1}^l \{ e^{iKU_{\zeta}\Gamma_{1,
 \zeta}\Gamma_{2m,\zeta}}\prod_{\rho=1}^{m-1}e^{iK\Gamma_{2\rho,\zeta }
 \Gamma_{2\rho+1, \zeta}} \} .
 \end{eqnarray}
 With the identity,
 \begin{equation}
 \big(U_{\zeta}\Gamma_{1,\zeta}\Gamma_{2m,\zeta} \big)^2=U_{\zeta}
 \Gamma_{1,\zeta}\Gamma_{2m,\zeta}U_{\zeta}\Gamma_{1,\zeta}
 \Gamma_{2m,\zeta} =-I\;,
 \end{equation}
  $\mathcal{V}_2$ can be rewritten as
 \begin{equation}
 \mathcal{V}_2=\prod_{\zeta=1}^l \{ \Big[ {1\over2}
 (I+U_{\zeta})e^{iK\Gamma_{1,\zeta}\Gamma_{2m,\zeta}}+{1 \over
 2} (I-U_{\zeta})e^{-iK\Gamma_{1,\zeta}\Gamma_{2m,\zeta}}
 \Big]\prod_{\rho=1}^{m-1}e^{iK\Gamma_{2\rho,
 \zeta}\Gamma_{2\rho+1, \zeta}} \} .
 \end{equation}
 Since $U_{\zeta}$ commutes with $\Gamma_{2\rho ,
 \zeta}\Gamma_{2\rho+1, \zeta}$, the projection operators ${1\over
 2}\;(I\;\pm\;U_{\zeta})$
 project $\mathcal{V}_2$ into $2^l$ pieces. For
 \begin{equation}
 \mathcal{V}_3 =\prod_{\rho=1}^m \Big( e^{KZ_{\rho, l}Z_{\rho, 1}}
 \prod_{\zeta=1}^{l-1} e^{KZ_{\rho, \zeta}Z_{\rho, \zeta +1}} \Big)
 \;,
 \end{equation}
 the situation seems more complicated.
 \begin{eqnarray}
 Z_{\rho, \zeta} Z_{\rho, \zeta+1} &=& \big( Z_{\rho, \zeta} Z_{\rho
 +1, \zeta} \big) \big( Z_{\rho +1, \zeta}Z_{\rho+2, \zeta} \big)
 \cdots \big( Z_{m,\zeta}Z_{1,\zeta+1} \big) \big(
 Z_{1,\zeta+1}Z_{2,\zeta+1} \big) \nonumber \\ & & \cdots \big(Z_{\rho-1, \zeta+1}
 Z_{\rho, \zeta+1 }\big) ,
  \\ &=& \big( -i\Gamma_{2\rho+1,\zeta} \Gamma_{2\rho, \zeta} \big)
 \big( -i \Gamma_{2\rho+3, \zeta}\Gamma_{2\rho+2, \zeta} \big)
 \cdots \nonumber \\ & & \big( -i\Gamma_{1, \zeta +1}\Gamma_{2m, \zeta}\big) \cdots
 \big( -i\Gamma_{2\rho-1, \zeta+1}\Gamma_{2\rho-2, \zeta+1} \big),
  \\ &=& i^m \Gamma_{2\rho, \zeta}\Gamma_{2\rho+1,
 \zeta}\Gamma_{2\rho+2, \zeta} \cdots \Gamma_{2m, \zeta}
 \Gamma_{1,\zeta+1} \cdots \Gamma_{2\rho-1,\zeta+1},
  \\ &=& i^m \Gamma_{2\rho, \zeta} \big( \Gamma _{2\rho+1,
 \zeta}\cdots \Gamma_{2\rho-2, \zeta+1} \big)
 \Gamma_{2\rho-1, \zeta+1},
  \\ &=& +i \Gamma_{2\rho, \zeta} W_{2\rho+1,\zeta}\Gamma_{2\rho-1,
 \zeta+1} ,
  \\ &=& iW_{2\rho+1, \zeta}\Gamma_{2\rho, \zeta}\Gamma_{2\rho-1,
 \zeta+1}  .
 \end{eqnarray}
$W_{2\rho+1,\zeta}$ is defined as
\begin{eqnarray}
W_{2\rho+1,\zeta} &=&
i^{m-1}\Gamma_{2\rho+1,\zeta}\Gamma_{2\rho+2, \zeta} \cdots
\Gamma_{2\rho-3, \zeta+1} \Gamma_{2\rho-2, \zeta+1}
 \\ \nonumber \\ &=& I \otimes \cdots I \otimes \underbrace{ \overbrace{\sigma_x}^{(\rho+1,\zeta)}
\otimes \cdots \otimes \overbrace{\sigma_x}^{(\rho-1, \zeta+1)}
}_{m-1\; times\; of\; \sigma_x}\otimes I \cdots \otimes I.
\end{eqnarray}
$W_{2\rho+1, \zeta}$ has the property that it anticommutes with
$\Gamma_{\mu , \alpha} $  inside the region that the integral
coordinates $(\mu ,\alpha )$ from ($2\rho+1, \zeta$) to (
$2\rho-2, \zeta +1$ ), whereas it commutes with $\Gamma_{\mu ,
\alpha } $ outside of that region.
\begin{eqnarray}
Z_{\rho, l}Z_{\rho, 1} &=& (Z_{\rho, l}Z_{\rho+1,l}) \cdots
(Z_{m,l}Z_{1,1})(Z_{1,1}Z_{2,1}) \cdots (Z_{\rho-1,1}Z_{\rho, 1})
 \\ &=& -i^m U \Gamma_{2\rho, l}\Gamma_{2\rho+1, l} \cdots
\Gamma_{2m,l} \Gamma_{1,1}\Gamma_{2,1} \cdots \Gamma_{2\rho-1, 1}
 \\ &=& +i UW_{2\rho+1,l}\Gamma_{2\rho-1,1}\Gamma_{2\rho,l}.
\end{eqnarray}
\begin{eqnarray}
W_{2\rho+1, l} &=& i^{m-1} \Gamma_{2\rho+1, l}
\Gamma_{2\rho+2,l}\cdots \Gamma_{2m,l} \Gamma_{1,1} \cdots
\Gamma_{2\rho-2, 1}  \\ &=& \underbrace{\sigma_x}_{(1,1)} \otimes
\cdots \otimes \underbrace{\sigma_x}_{(\rho-1, 1)} \otimes I
\otimes \cdots \otimes I \otimes \underbrace{\sigma_x}_{(\rho+1,
l)} \otimes \cdots \otimes \underbrace{\sigma_x}_{(m,l)} .
\end{eqnarray}
Then we have
\begin{equation}
\mathcal{V}_3 =\prod_{\rho=1}^m \{
e^{+iKUW_{2\rho+1,l}\Gamma_{2\rho-1,1}\Gamma_{2\rho,l}}
\prod_{\zeta=1}^{l-1}e^{iKW_{2\rho+1,\zeta}\Gamma_{2\rho,\zeta}
\Gamma_{2\rho-1, \zeta+1}}\}.
\end{equation}

A remarkable observation of Kaufman is that decomposing
$\mathcal{V}$ of the 2D Ising model into the product of  factors
like $e^{{\theta \over 2}\Gamma \Gamma}$, which is interpreted as
a two-dimensional rotation with rotation angle $\theta $ in the
direct product space. We follow this spirit and decompose the
factors into several 2D rotations,
\begin{eqnarray}
 & & e^{+iKUW_{2\rho+1,l}\Gamma_{2\rho-1,1}\Gamma_{2\rho,l}}
 \nonumber \\ &=&
{1\over2} (I+U)
e^{+iKW_{2\rho+1,l}\Gamma_{2\rho-1,1}\Gamma_{2\rho,l}}+{1\over2}
(I-U)e^{-iKW_{2\rho+1,l}\Gamma_{2\rho-1, 1}\Gamma_{2\rho,l}}
\\ \nonumber \\ &=& {1\over2} (I+U) \{\; {1\over2}
(I+W_{2\rho+1,l})e^{+iK\Gamma_{2\rho-1,1}\Gamma_{2\rho,l}}+{1\over2}
(I-W_{2\rho+1,l})e^{-iK\Gamma_{2\rho-1,1}\Gamma_{2\rho,l}} \;\}\;
 \nonumber \\ & & +{1\over2} (I-U) \;\{ {1\over2}
(I+W_{2\rho+1,l})e^{-iK\Gamma_{2\rho-1,1}\Gamma_{2\rho,l}}+{1\over2}
(I-W_{2\rho+1,l})e^{+iK\Gamma_{2\rho-1,1}\Gamma_{2\rho,l}}\;\}
\\ \nonumber \\ & &
e^{iKW_{2\rho+1,\zeta}\Gamma_{2\rho,\zeta}\Gamma_{2\rho-1,\zeta+1}}
\nonumber \\ &=& {1\over2}
(I+W_{2\rho+1,\zeta})e^{iK\Gamma_{2\rho,\zeta}\Gamma_{2\rho-1,\zeta+1}}
+{1\over2}
(I-W_{2\rho+1,\zeta})e^{-iK\Gamma_{2\rho,\zeta}\Gamma_{2\rho-1,
\zeta+1}}.
\end{eqnarray}
In the 2D Ising model the $\mathcal{V}$ are decomposed into 2
pieces. This is not so simple in the 3D Ising model. Due to the
projection operators $1\over2$ (I $\pm$ U), $1\over2$ (I $\pm$
$U_\zeta$), $1\over2$(I $\pm$ W), $\mathcal{V}$ has $2^l \times (2
\times 2 ^{ml})$ pieces. Only one piece will produce the largest
eigenvalue, which dominates the value of partition function.

Since $W_{2\rho+1,\zeta}$ may commute or anticommute with
$\Gamma_{\mu,\alpha}$, we have to check the commutation relations
between the projection operators and the product of the factors
like $e^{{1\over 2}\theta\Gamma \Gamma}$, since the dogma states
that two matrices are simultaneously diagonized if and only if
these two matrices commute. $\mathcal{V}$ comprises the product of
projection operators and a lot of $e^{{1\over 2}\theta\Gamma
\Gamma }$. So if the commutation relations are not valid, the
whole scheme may break down.  Commuting ${1\over2} (I \pm U)$ with
the product of all factors like $e^{{1\over 2}\theta\Gamma
\Gamma}$ does not give any trouble. Any single projection operator
${1\over2}(I\pm U_{\zeta})$ or ${1\over2}(I \pm W)$ may do not
commute with some $e^{{1\over 2}\theta \Gamma \Gamma}$. However,
fortunately, the product of all the projections in  any one piece
of $\mathcal{V}$, e.g.
\begin{equation}
P = {1\over2} (I+U) \Big( \prod_{\zeta'=1}^l {1\over2}
(I+U_{\zeta'}) \Big) \Big( \prod_{\rho=1}^m   \prod_{\zeta=1}^{l}
{1\over2} (I+W_{2\rho-1, \zeta}) \Big),
\end{equation}
do commute with any $e^{{1\over 2}\theta\Gamma \Gamma}$, because
$e^{{1\over 2}\theta\Gamma \Gamma}$ passes through P will change
 the sign of ${1\over 2} \theta \Gamma \Gamma $ even number times
and it does not change sign eventually.

 How to choose the proper
piece and obtain the largest eigenvalue? We have no answer. A
possible rule may be followed  that all  eigenvalue equations
should reduce to one equation. We will show this point in the next
section. This is a beautiful feature in the 2D Ising model. Let us
consider one possible piece of $\mathcal{V} $,
\begin{eqnarray}
\widetilde{\mathcal{V}} &=& (2 sinh 2K )^{ml/2} \;
\widetilde{\mathcal{V}}^{'} ,
\\ \nonumber \\
 \widetilde{\mathcal{V}}^{'} &=& \{ \prod_{\rho=1}^m
e^{+iK\Gamma_{2\rho-1,1}\Gamma_{2\rho,l}} \prod_{\zeta=1}^{l-1}
e^{iK\Gamma_{2\rho, \zeta}\Gamma_{2\rho-1, \zeta+1}}\} \nonumber
\\ & & \{ \prod_{\zeta'=1}^l
e^{iK\Gamma_{1,\zeta'}\Gamma_{2m,\zeta'}} \prod_{\rho'=1}^{m-1}
e^{iK\Gamma_{2\rho,\zeta}\Gamma_{2\rho+1, \zeta}} \} \nonumber \\
& & \{ \prod_{\zeta^{''}=1}^l \prod_{\rho^{''}=1}^m
e^{ik^{*}\Gamma_{2\rho^{''}-1,\zeta^{''}}\Gamma_{2\rho^{''},\zeta^{''}
}} \} .
\end{eqnarray}
In essence, $\widetilde{\mathcal{V} }$ includes the repetition $l$
times of the same rotations as in the 2D Ising model, appearing in
the second and third brackets of (79), and the new
 rotations in the first bracket of (79), relating to the
third dimensional coupling beyond the 2D Ising model.
\section{Eigenvalue Equations}
The  rotation operator in the spinor representation,
\begin{equation}
S_{\lambda\sigma}(\theta) = e^{{1\over2} \theta \Gamma_{\lambda}
\Gamma_{\sigma}}\;\;\;\;\;\;\;\;\;\; ( \lambda \neq \sigma) ,
\end{equation}
has a one-to-one correspondence to the 2D rotational matrix
$\omega(\lambda\sigma\mid \theta)$ of the $\Gamma$ matrix.
\begin{eqnarray}
S^{-1}_{\lambda\sigma}(\theta) \Gamma_{\alpha}
S_{\lambda\sigma}(\theta) &=& \sum_{\kappa} \omega (\lambda \sigma
\mid \theta )_{\alpha\kappa}\Gamma_{\kappa} ,
\\S^{-1}_{\lambda\sigma}(\theta)\Gamma_{\lambda}S_{\lambda\sigma}(\theta)
&=& \Gamma_{\lambda} \cos{\theta} +\Gamma_{\sigma} \sin{\theta} ,
\;\;\;\;\;\;\;\;\;\; (\lambda\neq\sigma),
\\ S^{-1}_{\lambda\sigma}(\theta)\Gamma_{\sigma}S_{\lambda\sigma}(\theta)
&=& -\Gamma_{\lambda}\sin{\theta} + \Gamma_{\sigma} \cos{\theta} ,
\;\;\;\;\;\;\;\;\;\;\;(\sigma\neq\lambda),
\\
S^{-1}_{\lambda\sigma}(\theta)\Gamma_{\alpha}S_{\lambda\sigma}(\theta)
&=& \Gamma_{\alpha} ,\;\;\;\;\;\;\;\;\;\;(\alpha\neq\lambda,
\alpha\neq\sigma).
\end{eqnarray}
$S_{\lambda\sigma}$($\theta$) is a 2D rotations in the direct
product space. The rotations, $e^{\pm {\theta \over 2}\Gamma
\Gamma}$, have different rotational angles, $+\theta$ and
$-\theta$. We mean they have different directions of rotations.
The eigenvalues of the rotational matrix $\omega $ are 1 with
$(2ml-2)$-fold degeneracies and $e^{\pm i\theta}$ two
nondegenerate eigenvalues, whereas the eigenvalues of $S_{\lambda
\sigma}$ in the spinor representation  are $e^{\pm i{\theta
\over2}}$  each with $2^{ml-1}$ -fold degeneracies.

The correspondence with $\widetilde{\mathcal{V}}$ is
\begin{eqnarray}
\widetilde{\omega} &=& (2 sinh 2K )^{ml/2} \;
\widetilde{\omega}^{'} \\ \nonumber \\ \widetilde{\omega}^{'} &=&
\{ \prod_{\rho=1}^m \omega (2\rho-1,1 ; 2\rho,l \mid +2iK)
\prod_{\zeta=1}^{l-1} \omega (2\rho,\zeta;2\rho-1, \zeta+1 \mid
2iK) \} \nonumber
\\ & & \{ \prod_{\zeta^{'}=1}^l \omega (1,\zeta^{'};2m,\zeta^{'}
\mid 2iK) \prod_{\rho^{'}=1}^{m-1} \omega (2\rho, \zeta ;2\rho +1,
\zeta \mid 2iK ) \} \nonumber
\\ & & \{ \prod_{\zeta^{''}=1}^l \prod_{\rho^{''}=1}^m \omega
(2\rho^{''} -1, \zeta^{''}; 2\rho^{''}, \zeta^{''} \mid 2iK^{*} )
\}
\\ \nonumber \\ &=& \omega_3 \omega_2 \omega_1 . \\ \nonumber
\\  \omega_{1}^{1\over2}\omega_{2}\omega_{1}^{1\over2} &=&  \left(
\begin{array}{lll} \displaystyle{\begin{array}{cc} \Omega
&   \\   & \Omega
\end{array}} & \displaystyle{\begin{array}{cc}   &   \\   &
\end{array}} & \displaystyle{\begin{array}{c}
\end{array}}
  \\ \displaystyle{\begin{array}{cc}   &   \\   &
\end{array}} & \displaystyle{\begin{array}{cc} \ddots &   \\   &
\ddots
\end{array}} & \displaystyle{\begin{array}{cc}  &   \\   &
\end{array}} \\ \displaystyle{\begin{array}{c}
\end{array}} & \displaystyle{\begin{array}{cc}   &   \\   &
\end{array}} & \displaystyle{\begin{array}{cc} \Omega &   \\   &
\Omega
\end{array}} \end{array} \right)_{2ml\times 2ml}  ,
\\ \Omega &=& \left(
\begin{array}{c} A \\ B^{\dagger} \\ 0 \\    \\ 0 \\ -B \end{array}
 \begin{array}{c} B \\ A \\ B^{\dagger} \\   \\ \cdots \\ 0
\end{array} \begin{array}{c} 0 \\ B \\ A \\     \\  0 \\ \cdots
\end{array} \begin{array}{c} \cdots \\ 0 \\ B \\ \ddots \\ B^{\dagger} \\
0 \end{array} \begin{array}{c} 0 \\ \cdots \\ 0 \\   \\ A \\
B^{\dagger}
\end{array} \begin{array}{c} -B^{\dagger} \\ 0 \\ \cdots \\    \\ B \\ A
\end{array} \right)_{2m\times 2m}  ,
\end{eqnarray}
\begin{eqnarray}
A  &=& \left( \begin{array}{cc} c^{*}c & is^{*}c \\ -is^{*}c &
c^{*}c \end{array} \right) ,\;\;\;\; B=\left( \begin{array}{cc}
-{1\over2} & -is({{-1+c^{*}}\over2})
\\ is({{1+c^{*}}\over2}) &
-{1\over2}
\end{array} \right),\\ B^{\dagger} &=& \left( \begin{array}{cc}
-{1\over2} & -is({{1+c^{*}}\over2})
\\ is({{-1+c^{*}}\over2}) & -{1\over2}
\end{array} \right),
\end{eqnarray}
\begin{equation}
s\equiv \sinh 2K, \;\;c\equiv \cosh 2K, \;\; s^{*} \equiv \sinh
2K^{*},\;\; c^{*} \equiv \sinh 2K^{*}.
\end{equation}
The matrix $\Omega$ is just the same matrix considered in the 2D
Ising model. $\omega_1^{1\over2}\omega_2 \omega_1^{1\over2}$ is in
symmetric form such that its eigenvalue equations are much more
easier to handle.
\begin{eqnarray}
\omega_3 &\equiv& \prod_{\rho=1}^m \omega \Big( 2\rho-1, 1;
2\rho,l \mid 2iK \Big) \prod_{\zeta=1}^{l-1} \omega \Big( 2\rho ,
\zeta; 2\rho-1, \zeta +1 \mid 2iK \Big) \\ & = & \left(
\begin{array}{c} \mathcal{A} \\ \mathcal{B}^{\dagger} \\ 0 \\   \\ 0 \\
\mathcal{-B}
\end{array} \begin{array}{c} \mathcal{B} \\ \mathcal{A} \\
\mathcal{B}^{\dagger} \\    \\ \cdots \\ 0 \end{array}
\begin{array}{c} 0
\\ \mathcal{B} \\ \mathcal{A} \\    \\ 0 \\ \cdots \end{array}
\begin{array}{c} \cdots \\ 0 \\ \mathcal{B} \\ \ddots \\
\mathcal{B}^{\dagger} \\ 0 \end{array} \begin{array}{c} 0 \\
\cdots
\\ 0
\\   \\ \mathcal{A} \\ \mathcal{B}^{\dagger} \end{array}
\begin{array}{c} \mathcal{-B}^{\dagger} \\ 0 \\ \cdots \\    \\
\mathcal{B} \\ \mathcal{A} \end{array} \right)_{2ml\times 2ml },
\end{eqnarray}
\begin{eqnarray}
\mathcal{A} &=& \left( \begin{array}{c} c \\    \\    \\
 \end{array} \begin{array}{c}     \\ c \\   \\    \end{array}
\begin{array}{c}   \\   \\ \ddots \\    \end{array}
\begin{array}{c}
  \\   \\   \\ c  \end{array} \right)_{2m\times 2m}  ,\;\;\;\;
\mathcal{B}^{\dagger} = \left( \begin{array}{c} 0 \\ 0 \\ 0 \\
\vdots
\\
  \\   \\   \\  \end{array} \begin{array}{c} -is \\ 0 \\ 0 \\ \vdots
  \\  \\   \\   \\  \end{array} \begin{array}{c} 0 \\ 0 \\ 0 \\
  \vdots \\   \\   \\   \\   \end{array} \begin{array}{c} \cdots \\
  \cdots \\ -is \\ 0 \\ \vdots \\   \\   \\   \end{array}
  \begin{array}{c}   \\   \\ 0 \\ 0 \\  \\   \\   \\   \end{array}
  \begin{array}{c}   \\   \\ \cdots \\ \cdots \\ \ddots \\   \\   \\
  \end{array} \begin{array}{c}  \\   \\   \\   \\   \\   \\   \\
  \end{array} \begin{array}{c}   \\   \\   \\   \\   \\   \\ 0 \\ 0
  \end{array} \begin{array}{c}   \\   \\   \\   \\   \\   \\ -is \\ 0
  \end{array}
  \right)_{2m\times 2m},
  \\  \mathcal{B} &=&  \left( \begin{array}{c} 0 \\ is \\ 0 \\
  \vdots \\   \\   \\   \\   \end{array}  \begin{array}{c} 0 \\ 0 \\ 0 \\
  \vdots \\   \\   \\   \\    \end{array} \begin{array}{c} \cdots \\
   0 \\ 0 \\ is \\   \\   \\   \\   \end{array} \begin{array}{c} \cdots
    \\ \cdots \\ 0 \\ 0 \\   \\   \\   \\   \end{array} \begin{array}{c}
   \\   \\ 0 \\ 0 \\ \ddots \\   \\   \\   \end{array}
   \begin{array}{c}   \\   \\ \cdots \\ \cdots \\   \\   \\   \\
\end{array} \begin{array}{c}  \\   \\  \\   \\   \\   \\ 0 \\ is \end{array}
\begin{array}{c}  \\   \\   \\   \\   \\   \\ 0 \\ 0
\end{array}\right)_{2m\times 2m}.
\end{eqnarray}
Now we proceed to solve the eigenvalue equation
\begin{equation}
\widetilde{\omega} \Psi = \lambda \Psi,
\end{equation}
where $\widetilde{\omega}= (2 sinh2K )^{ml/2}\omega_3 \big[
\omega_1^{1/2} \omega_2 \omega_1^{1/2} \big] $,
\begin{equation}
\Psi = \left( \begin{array}{c} z\;\psi_0 \\ z^2\; \psi_0 \\ z^3
\;\psi_0  \\ \vdots \\ z^l\; \psi_0 \end{array}
\right)\;,\;\;\;\;\; \psi_0 = \left( \begin{array}{c} y\; u \\
y^2\; u \\ y^3\; u \\ \vdots \\ y^m\; u \end{array} \right)\;,
\;\;\;\;\; u=\left(
\begin{array}{c} u_1 \\ u_2 \end{array} \right) .
\end{equation}
By imposing the constraint,
\begin{equation}
z^l = \; -1\;, \;\;\;\;\; or \;\; z= e^{i{{\pi t_2}\over
l}}\;\;\;\;\; (\;t_2=1,3,5 \cdots\;),
\end{equation}
one reduces eigenvalue equation (97) to
\begin{equation}
 \big( \mathcal{A} +z \; \mathcal{B} +z^{-1} \;
\mathcal{B}^{\dagger} \; \big) \;\Omega \; \psi_0 \;=\; \lambda \;
\psi_0\;.
\end{equation}
Further, imposing the constraint,
\begin{equation}
y^m\;=\;-1\;,\;\;\;\;\;  or \;\; y\;=\;e^{i{{\pi t_1}\over
m}}\;\;\;(\;t_1\;=\; 1,3,5 \cdots\;),
\end{equation}
(100) is reduced to
\begin{equation}
\mathcal{D} \big(\;A\;+y\;B\;+y^{-1} B^{\dagger} \big) \;
u\;=\;\lambda u,
\end{equation}
where
\begin{equation}
\mathcal{D} = \left( \begin{array}{cc}  c  &  -iz^{-1}s  \\ iz\;s
& c
\end{array} \right)\;,
\end{equation}

\begin{eqnarray}
 & & A+yB+y^{-1}B^{\dagger} \\ &=& \left(\begin{array}{cc}
c^{*}c-\cos{{\pi t_1}\over m} & is^{*}c - i( {{-1+c^{*}}\over 2})
s e^{{i\pi t_1}\over m} -i( {{1+c^{*}}\over 2}) s e^{-i{{\pi
t_1}\over m}} \\ -is^{*}c +i( {{1+c^{*}}\over2})s e^{i{{\pi
t_1}\over m}} + i( {{-1+c^{*}} \over 2}) s e^{-i{{\pi t_1}\over
m}} & c^{*} c - \cos{{\pi t_1}\over m}
\end{array}\right).\nonumber
\end{eqnarray}
The eigenvalue equation (102) is further reduced to
\begin{equation}
\lambda^2 \;-\; 2 \cosh{\gamma}\;\; \lambda\; +\; 1 \;=\;0\; .
\end{equation}
$\lambda =\exp \big( \pm \gamma \big) $ is the solution of
$\lambda\;$.
 $\gamma $ is determined by
\begin{equation}
\cosh{\gamma} \;=\; {c^3\over s} - c(\cos{\theta_1}
+\cos{\theta_2} ) + s c\; \cos{\theta_1} \cos{\theta_2}+
s^2\sin{\theta_1} \sin{\theta_2},
\end{equation}
where two continuous variables, $\theta_1$, $\theta_2$, are
obtained by taking the
 thermodynamic limit, $m,l \rightarrow \infty$, $ {{\pi t_1}\over
m}\rightarrow \theta_1$ , $ {{\pi t_2 }\over l }\rightarrow
\theta_2 $.

The partition function of $\mathcal{V}$ is
\begin{equation}
 \mathcal{Z} \sim \lambda^N .
\end{equation}
The free energy per site under the thermodynamic limit is
\begin{eqnarray}
f &=& -{1\over \beta } \lim_{N\rightarrow \infty} N^{-1} ln
\mathcal{Z} \\ &=& -{1\over \beta} \ln\; (2\sinh 2K ) ^{1/2} -
{1\over 8{\pi}^2 \beta } \int_0 ^{2\pi} \int_0 ^{2\pi} \; \gamma
\; (\theta_1, \theta_2 \; ) d \theta_1 d \theta_2 ,
\end{eqnarray}
where $\beta \equiv  {1\over T}.$ For convenience, $J\equiv 1$,
the internal energy per site is
\begin{eqnarray}
u &=& {\partial \over {\partial {\beta}}}(\beta f ) \\ &=&  -
\coth {2K} - {1\over {8\pi^2}} \int_0^{2\pi} \int_0^{2\pi}
{\partial \gamma \over {\partial K }} d\theta_1 d\theta_2 .
\end{eqnarray}
\section{High Temperature Limit}
Let us expand $u$ in terms of $x$ at high temperature limit, $x$
small, and compare the series of $u$ obtained from the computer
graphic method.\cite{DG}
 \begin{equation}
 x  \equiv \; \tanh{K} \;,\;\;\; c = \; \cosh{2K} = \;
 {{1+x^2}\over{1-x^2}},\;\;\; s=\; \sinh{2K} =\;
 {{2x}\over{1-x^2}}\;,
 \end{equation}
\begin{eqnarray}
\cosh{\gamma }  &=& \;{{\big( 1+x^2 \big)^3}\over {\big(
1-x^2\big) ^{2}2x}}- {{1+x^2}\over {1-x^2}} (\; \cos {\theta_1}
+\cos {\theta_2}  )\nonumber
\\ \nonumber \\  & & + {{2x\big( 1+x^2\big) }\over{(1-x^2)^2}} \cos{\theta_1}
\cos{\theta_2} + {{4x^2}\over{\big( 1-x^2\big) ^2}} \sin{\theta_1}
\cos{\theta_2} ,
\\ \nonumber \\ {{\partial \gamma } \over {\partial K}}  &=& \; 2
(\sinh{\gamma} )^{-1} \;\{ \; -{{(1+x^2)^2 (1-10x^2+x^4)}\over
{(1-x^2)^2 4x^2 }}  - \; {{2x} \over {1-x^2} } (\cos{\theta_1} +
\cos{\theta_2} ) \nonumber \\ \nonumber \\ \;\;\;\;\;\;\;\;\;\; &
& + {{1+6x^2 +x^4}\over {(1-x^2)^2}} \cos{\theta_1} \cos{\theta_2}
+\; {{4x(1+x^2)}\over {(1-x^2)^2}} \sin{\theta_1} \sin{\theta_2}
\; \} .
\end{eqnarray}
With the help of  computer program Maple V Release 5.1, we get our
result
\begin{equation}
u\; = \;-3x-8x^3-28x^5-132x^7-832x^9+O(x^{11}).
\end{equation}
Comparing the result obtained by the computer graphic method, the
series expansion of $u$ for a simple cubic lattice can be
transformed from the partition function,
\begin{equation}
\mathcal{Z}^{1\over N} = 2cosh^3 K \;  \big( 1+\; 3x^4 +\; 22x^6
+\; 187.5x^8
 +\; 1980x^{10} +\; O(x^{12}) \big),
 \end{equation}
so
\begin{eqnarray}
u &=&\; -{{\partial}\over{\partial{K}}}  \ln  \mathcal{Z}^{1\over
N}
\\ \nonumber \\ &=&
\;-3x-12x^3- 120x^5-1332x^7-17676x^9 +O(x^{11}).
\end{eqnarray}
(115 ) and (118) have very similar structures. The coefficient of
$x^{-1} $ does not exist  in our result (115) though it exists
naively in $coth 2K$ of (111). The first nonvanishing term $-3x$
of (115) and the vanishing of all coefficients of even powers of
$x$ are the same as (118 ). All terms, up to $O(x^9)$, with minus
signs in (115) are also the  same as (118). The reason why (115)
and (118) do not have the same first three or four coefficients
may come from the fact that we cannot decipher precisely the
largest eigenvalue of $\mathcal{V}$ . On the other hand, if we
choose different directions for the 2D rotations, the eigenvalue
equations may not be reduced to only one equation. This is a
dilemma.

 The correct
relation of $\cosh{\gamma}$ implies the correct thermodynamic
quantities of the 3D Ising model. To guess the correct relation as
(106 ) may be a promising way for finding the right resolution of
the 3D Ising model. For example, trying to repair (106), if we
multiply the matrices $\mathcal{A}$, $\mathcal{B}$ and
$\mathcal{B}^{\dagger}$, or $\omega_3$, with a correcting factor
$\Phi (x)$, then we have
\begin{equation}
\cosh{\gamma} = {{c^3}\over s }- c \cos{\theta_1}+ \Phi (-c
\cos{\theta_2}+sc\cos{\theta_1}\cos{\theta_2}+ s^2
\sin{\theta_1}\sin{\theta_2}) .
\end{equation}
With the help of Maple V, setting
\begin{equation}
\Phi (x) =1- x^2 -{27\over 2}x^4 -{249\over 2}x^6 -{12325\over
8}x^8-O(x^{10} ) ,
\end{equation}
 then the series expansion of $u$ has the same result as (118). $\Phi(x)$ may be explained
as, something like, a weighting function for different directions
of 2D rotations.
\section{Acknowledgements}
One of authors, S.L. Lou, would like to thank H.C. Lee, Friday Lin
and C.Y. Lin for many helpful discussions about four years ago.


\newpage
\begin{thebibliography}  {xx}
\bibitem{On} L.Onsager Phys. Rev. {\bf 65}
             (1944) 117.
\bibitem{DG} C.Domb, M.S.Green(eds.), Phase Transitions and Critical
             phenomeno, (Academic Press).{\bf Vol.3} (1974).
\bibitem{Ka} B.Kaufman,Phys.Rev. {\bf 76}, (1944) 1232.
\bibitem{ZJ} J.Zinn-Justin, J.Physique {\bf 42} (1981) 783.
\bibitem{LF} A.J.Liu and M.E.Fisher, Physica {\bf A156} (1989)35.
\bibitem{GE} A.J.Guttmann and I.G.Enting, J.Phy. {\bf A27} (1994)8007.
\bibitem{NM} G.F.Newell and E.W.Montroll, Rev.Mod.Phys. {\bf 25}(1953)353.
\bibitem{RE} G.S.Rushbrooke, andJ.Eve, J.Math.Phys. {\bf 3} (1962)185.
\bibitem{IT} C.Itzykson, Nucl. Phys. {\bf 210} (1982)477.
Itoi, Phys. Lett. {\bf 73}(1994)3335. M.Asano, C.Itoi and
S.Kojima, Nucl. Phys.{\bf B448} (1995)533. Shi He, Xu Yi-Chan and
Hao Bai-Lin Acta Physica Sinica {\bf 27} (1978)47 ; { \bf
29}(1980)1564; {\bf 30}(1981)1225. M.E.Fisher, J.Phys. {\bf
A28}(1995)6323. R. Guida and Zinn-Justin, Nucl. Phys. {\bf
B489}(1997)626. M.Caselle and M.Hasenbusch, J.Phys. {\bf
A30}(1997)4963.
\end{thebibliography}
\end{document}